\documentstyle[preprint,aps]{revtex}

\begin{document}

\draft

\title{An updated analysis of NN elastic scattering data to 1.6~GeV}
\author{Richard A. Arndt, Igor I. Strakovsky$^\dagger$ and Ron L. Workman}
\address{Department of Physics, Virginia Polytechnic Institute and State
University, Blacksburg, VA 24061}

\date{\today}
\maketitle

\begin{abstract}
An energy-dependent and set of single-energy partial-wave analyses of $NN$
elastic scattering data have been completed.  The fit to 1.6~GeV has been
supplemented with a low-energy analysis to 400 MeV. Using the low-energy
fit, we study the sensitivity of our analysis
to the choice of $\pi NN$ coupling constant. We also comment on the
possibility of fitting $np$ data alone. These results are
compared with those found in the recent Nijmegen analyses.

\end{abstract}
\vskip 0.3in
\pacs{PACS numbers: 11.80.Et, 13.75.Cs, 25.40.Cm, 25.40.Dn}

\narrowtext
\section{Introduction}
\label{sec:intro}

This analysis of elastic nucleon-nucleon scattering data updates the
content of Ref.\cite{ar92}.
In the intervening period, a substantial amount of new
$np$ data has been accumulated. These additions to the database are
the subject of Section~II.
In Section~III, we give the results of our analyses and compare with
our previous solutions~\cite{ar92,ar87,ar83}
and those produced by the Nijmegen group~\cite{ni93}.

The Nijmegen group has continued to analyze data in the low-energy region and
now~\cite{fb14} claims the ability to fit both the I=0 and I=1 phases using
$np$ data alone. In order to explore
the low-energy region more closely, an analysis to 400 MeV (VZ40) was
carried out. Using VZ40 we considered the sensitivity of our
fits to the choice of pion-nucleon coupling constant, and
carried out fits to the $pp$ and $np$ data separately.
We have also studied the effect of pruning high-$\chi^2$
data from the database. Our finding are summarized in Section~IV.

\section{The Database}
\label{sec:dat}

Our previous $NN$ scattering analyses\cite{ar92} were based on
11,880 $pp$ and 7572 $np$ data. In Ref.\cite{ar92} the $pp$
analysis extended up to a laboratory kinetic
energy of 1.6~GeV; the $np$ analysis was truncated at 1.1 GeV.
The present database is considerably larger due both to an expanded energy
range for the $np$ system (up to 1.3~GeV) and the addition of new data.
The distribution of recent (post-1991) $pp$ and $np$ data is given in Fig.~1.
The total database has doubled over the last decade (see Table~I).
New $np$ data, mainly produced by LAMPF and Saclay since 1991,
have resulted in a better
balance between $pp$ and $np$ datasets.  In fact,
the $np$ database has increased by a factor of 1.3 since 1991.  Unfortunately,
we cannot extend our analysis of the I~=~0 system up to
a nucleon kinetic energy
of 1.6~GeV, due to the lack of $np$ data between 1.3~GeV and 1.6~GeV.

Since most of the new data~[6-45] are from
high-intensity facilities, they have added weight
against the older data.  Most of new $pp$ data were produced between 500 and
800~MeV.  LAMPF, for instance, has produced differential cross
sections\cite{si93}, polarization variables P\cite{gl93}, and correlation
parameters A$_{zz}$, A$_{zx}$\cite{gl92}, and A$_{yy}$\cite{ho94}.
Excitation measurements of P were carried out at KEK\cite{ko94} for
37$\pm$2$^{\circ}$ between 491~MeV and 1600~MeV and Saclay\cite{be92} for
43$^{\circ}$ between 523~MeV and 708~MeV.

Most of new $np$ were either measured below 100~MeV or
between 350~MeV and 1100~MeV.
The sources of low energy data are TUNL, PSI, and Uppsala which gave
d$\sigma$/d$\Omega$\cite{ro92}; P\cite{sc91p}, \cite{br92}, \cite{we92};
A$_{zz}$\cite{ha91}; A$_{yy}$\cite{kl94p}; and D$_t$\cite{oc91}.  LAMPF
has completed a 10 year $np$ program, producing data for
d$\sigma$/d$\Omega$\cite{no93}; A$_t$, A'$_t$, R$_t$, R'$_t$\cite{mc91} and
\cite{mc92}; A$_{yy}$, A$_{xx}$, A$_{zx}$ and A$_{zz}$\cite{di92}, \cite{sh93},
and\cite{sp94p}; D$_t$ and P\cite{mc93} and \cite{gl93}.  A detailed study of
$np$ polarization quantities was carried out at Saclay, producing data for
P\cite{ba931},
\cite{ba932}; A$_{yy}$\cite{ba933}; A$_{zz}$\cite{ba934}; A$_{zx}$\cite{ba935};
A$_t$, N$_{0nkk}$, D$_{0s"0k}$, R$_t$, N$_{0nsk}$, D, and D$_t$\cite{ba941}.
Total $np$ cross sections in pure spin states were also measured\cite{bi91},
\cite{fo91}, \cite{be93p}, \cite{ha93}, \cite{wi93}, \cite{ba942}.

\section{Partial-Wave Analysis}
\label{sec:pwa}

As mentioned in the introduction, this analysis extended to 1.6 GeV, with
an $np$ component up to 1.3 GeV. The energy-dependent solution required
77 isovector and 44 isoscalar parameters. The solution (FA91) described in
Ref.\cite{ar92} had 123 free parameters. The present
energy-dependent solution gives a $\chi^2$/datum
of 22371/12838 for $pp$ data and 17516/10918 for $np$ data.
A comparison with several of our previous
solutions is given in Table~I. In addition to the energy-dependent analysis,
single-energy fits of the $pp$ and $np$ data were obtained up to 1.25 GeV.
Two further analyses of $pp$ data alone were added at 1.3 GeV and 1.6 GeV.
These are described in Table~II, where we list the number of
varied parameters in each single-energy fit and compare with the
$\chi^2$ found in the energy-dependent solution. These single-energy results
are plotted with uncertainties in Fig.~2.

The most significant changes to FA91~\cite{ar92} were made in the
parameterization of the S-waves and in the tuning of the deuteron pole
parameters. The solutions FA91 and SM94 differ little in
the isovector partial-waves; only
the isoscalar waves are plotted in Fig.~2. Here we have displayed both
SM94 and FA91 for the purpose of comparison.  Large variations
are seen in the $^3D_2$ partial-wave, and at low-energies in $\epsilon_1$.
In Fig.~3, some prominent partial-waves are plotted in an Argand
diagram~\cite{dbs}.

In order to ascertain that the full fit to 1.6~GeV (1.3~GeV for np)
was not seriously degraded at low energies, a 0$-$400 MeV fit was also
developed. The resultant solution, VZ40, used 26 I=1 and 27 I=0 variable
parameters to give a $\chi^2$/datum of 3098/2170(pp), and 4595/3367(np).
The global fit, SM94, produced, for the same energy range, a
$\chi^2$/datum of 3443/2170(pp) and 5290/3367(np).
We consider this quite reasonable
given that the number of variable parameters per datum is nearly twice
as large for VZ40 as it is for SM94.
A comparison of selected phases is given in Fig.~4. Here we have also
compared with the Nijmegen analysis~\cite{ni93}. Note that while substantial
differences are seen between the Nijmegen and SM94 results for the $^1P_1$
and  $^3P_0$ phases, the VZ40 and Nijmegen results are quite consistent. The
most noticeable disagreement is seen in $\epsilon_1$.

To illustrate the stability of our solution (either VZ40 or SM94) against
pruning of the database, we performed the following exercise with VZ40.
The dataset was first pruned by discarding all data with $\chi^2$
contributions greater than 9; this resulted in the removal of 74 data
(27 pp and 47 np) with a consequent decrease in $\chi^2$ of about 1000.
The solution was then searched and $\chi^2$ decreased by a mere 45.
When we further pruned data giving $\chi^2$ contributions in
excess of 7, 71 more points were removed with a reduction of 590 in
$\chi^2$. Further searching reduced $\chi^2$ by only 14. The resultant,
pruned fit gave a $\chi^2$/datum of 2397/2112(pp) and 3643/3280(np)
with virtually {\it no} detectable change in the resultant phases.
Our $\chi^2$ values are clearly dependent upon the existence of
poorly fitted data. However, the solution itself appears quite
insensitive to the removal of high-$\chi^2$ data.

In joint analyses of $pp$ and $np$ data, it is commonly assumed that the
I=1 phases are essentially determined by the $pp$ data. If I=1 phases
could be determined directly from the $np$ data, this would provide an
interesting check on charge independence. Until recently, this was not
possible. However, the Nijmegen group claims\cite{fb14}
to have succeeded in an
analysis of the $np$ data alone, and have compared their results to
those coming from analyses of the $pp$ data alone.  We have attempted this
using our VZ40 solution and find that it is indeed possible to fit the
$np$ data separately.

We first attempted to fit the $pp$ data alone, in
order to determine the effect of $np$ data on the I=1 phases. The $np$ data
were removed and the solution adjusted to best fit the $pp$ measurements. The
$\chi^2$ for the $pp$ data dropped from 3098 to 3083. This is what one
would expect if only the $pp$ data were significantly influencing the I=1
phases. More surprising was the effect of removing all $pp$ data from the
0$-$400 MeV database. A stable solution was found with $\chi^2$ changing
from 4595 to 4422 for the $np$ data. The small decrease in $\chi^2$ suggests
that charge independence is a reasonable assumption in joint analyses of
$np$ and $pp$ data.

Sensitivity to the pion-nucleon coupling, $g^2/ 4\pi$ was probed by mapping
$\chi^2$ versus $g^2/ 4\pi$ for the solution VZ40. The results are illustrated
in Fig.~5 where we have plotted the changed in $\chi^2$ for $pp$, $np$ and
combined data. The resulting parabola for combined data shows a consistency
with our chosen value (13.7), but with a rather weak sensitivity. We
do not consider this to be a reliable determination of $g^2/ 4\pi$
because it is dependent upon the particular way in which we
account for the one-pion-exchange in our representation.
In Fig.~5, H-waves and higher were treated in a
one-pion-exchange approximation. Purely for comparison purposes, we have
included in Fig.~5 the parabola which resulted from a $\chi^2$ mapping in
our pion-nucleon analysis\cite{ar94a}
to 2.1~GeV in the pion laboratory kinetic energy.
This analysis was based on more data (by a factor of 4) than were used in
the VZ40 fit, but the sensitivity is clearly much greater in our
pion-nucleon analysis.

\section{Results and Comparisons}
\label{sec:rs}

We have incorporated a large new set of $NN$ elastic scattering data into
our analyses. This set was mainly comprised of $np$ measurements,
and these produced noticeable changes in some isoscalar partial-waves.
The isovector waves remained fairly stable.
At low energies, $\epsilon_1$ changed significantly from the FA91 results.
Also at low energies, apart from  $\epsilon_1$,
comparisons between VZ40 and the Nijmegen results show good agreement.

In other tests with VZ40, we found that our solution was quite stable to
the removal of high-$\chi^2$ data. We also verified that the $np$ data
could be analyzed separately. The $\chi^2$ values for separate fits of the
$pp$ and $np$ data were not very different from results found in combined
analyses. We also demonstrated that a value for the $\pi NN$ coupling,
consistent with our $\pi N$ elastic scattering results, could be determined
from VZ40. We should emphasize that this was a consistency check and not
a determination of the coupling.

Some new TUNL measurements\cite{i1} of the P parameter for the np elastic
scattering at 8 and 12~MeV will soon be available.  While only a few
polarization quantities have been measured at medium energies, some new PSI
measurements\cite{i2} of RT and DT between 260 and 550~MeV and a few new
Indiana data\cite{i3} of P and AYY at 180~MeV will soon be available.

This reaction is incorporated into the SAID program\cite{tel}, which is
maintained at Virginia Tech.  Detailed information regarding the database,
partial-wave amplitudes and observables may be obtained either interactively,
through the SAID system (for those who have access to TELNET), or directly
from the authors.

\acknowledgments

The authors express their gratitude to R. Abegg, J. Ball, D.V. Bugg,
T. Bowyer, F. Bradamante, J.M. Durand, G. Glass,
L.G. Greeniaus, R. Henneck, G.A. Korolev, H. Klages, F. Lehar,
C. Lechanoine-LeLuc, M.W. McNaughton, D.F. Measday, W. Kretschmer,
V.N. Nikulin, W.K. Pitts, J. Rapaport, E. R\"ossle, H. Shimizu, H.M. Spinka,
W. Tornow, P. Truoel, S. Vigdor, R. Weidmann, W.S. Wilburn, and A. Yokosawa
for providing experimental data prior to publication or clarification of
information already published.  We thank H. von Geramb for a number of
useful communications regarding the deuteron parameters.
I.S. acknowledges the hospitality extended
by the Physics Department of Virginia Tech.  This work was supported in
part by the U.S.~Department of Energy Grant DE--FG05--88ER40454.


\newpage
{\Large\bf Figure captions}\\
\newcounter{fig}
\begin{list}{Figure \arabic{fig}.}
{\usecounter{fig}\setlength{\rightmargin}{\leftmargin}}
\item
{Energy-angle distribution of recent (post-1991) (a) $pp$ and (b) $np$ data.
$pp$ data are [observable (number of data)]: d$\sigma$/d$\Omega$~(81),
P~(383), D~(8), R~(6), A~(2), A$_{yy}$~(10), A$_{zz}$~(147), and
A$_{zx}$~(64). $np$ data: d$\sigma$/d$\Omega$~(221), P~(924), D~(30),
D$_t$~(128), A$_{yy}$~(389), A$_{xx}$~(159), A$_{zz}$~(415), A$_{zx}$~(425),
R$_t$~(103), R'$_t$~(80), A$_t$~(142), A'$_t$~(85), N$_{0nsk}$~(20),
D$_{0s"0k}$~(20), N$_{0nkk}$~(20), $\Delta \sigma _{T}$~(31),
$\Delta \sigma _{L}$~(23), and other~(5).  Total cross sections are
plotted at zero degrees.}
\item
{Isoscalar partial-wave amplitudes from 0 to 1.2~GeV.  Solid (dashed) curves
give the real (imaginary) parts of amplitudes corresponding
to the SM94 solution.  The real (imaginary) parts of single-energy solutions
are plotted as filled (open) circles. The previous FA91 solution
\cite{ar92} is plotted with long dash-dotted (real part) and short dash-dotted
(imaginary part) lines.  The dotted curve gives the value of
Im~T - T$^2$ - T$^{2}_{sf}$, where T$^{2}_{sf}$ is the spin-flip amplitude.
All amplitudes have been multiplied by a factor of 10$^3$ and are
dimensionless.}
\item
{Argand plot of the $NN$ partial-wave amplitudes $^1$D$_2$,
$^3$P$_2$, $^3$F$_3$, and $^1$G$_4$. (Compare
Figures~7 of references\cite{ar93} and \cite{ar94}).  The ``X" points denote
100~MeV steps.  All amplitudes have been multiplied by a factor of $10^3$
and are dimensionless.}
\item
{Phase--shift parameters from 0 to 400 MeV.  The SM94 and VZ40
solutions are plotted as solid and dash-dotted curves, respectively.
Single-energy solutions are given by filled circles.  A recent solution
from the Nijmegen group\cite{ni93} is plotted as a dashed curve.}
\item
{A plot of $\chi ^2$ versus g$^2$/4\,$\pi$. $\chi^2$ values are plotted as
deviations from the $pp$ and $np$ minima.
The open squares (triangles) give the
VZ40 results of $pp$ ($np$) data.  The black circles give the result of
a combined fit to both $pp$ and $np$ data. Solid lines drawn are to
guide the eye.  For the purpose of comparison, a $\chi^2$ map for
the recent FA93 $\pi$N solution\cite{ar94a} has been added
as a dashed curve.}
\end{list}

\newpage
\mediumtext
\vfill
\eject
Table~I. Comparison of present (SM94, VZ40) and previous (FA91, SM86, and
SP82) energy-dependent partial-wave analyses.  The $\chi ^2$ values for
the previous FA91, SM86, and SP82 solutions correspond to the published
results(\cite{ar92}-\cite{ar83}).
\vskip 10pt
\centerline{
\vbox{\offinterlineskip
\hrule
\hrule
\halign{\hfill#\hfill&\qquad\hfill#\hfill&\qquad\hfill#\hfill
&\qquad\hfill#\hfill&\qquad\hfill#\hfill&\qquad\hfill#\hfill\cr
\noalign{\vskip 6pt} %
Solution&Range~(MeV)&$\chi^2$/$pp$~data&Range~(MeV)&$\chi^2$/$np$~data&
Ref. \cr
\noalign{\vskip 6pt}
\noalign{\hrule}
\noalign{\vskip 10pt}
SM94 & $0 -1600$ &  22371/12838 & $0 -1300$ & 17516/10918 & Present \cr
\noalign{\vskip 6pt}
     &$(0  -400)$&   3443/2170  &$(0  -400)$&  5290/3367  & Present \cr
\noalign{\vskip 6pt}
VZ40 & $0  -400$ &   3098/2170  & $0  -400$ &  4595/3367  & Present \cr
\noalign{\vskip 6pt}
FA91 & $0 -1600$ &  20600/11880 & $0 -1100$ & 13711/7572  & \cite{ar92} \cr
\noalign{\vskip 6pt}
SM86 & $0 -1200$ &  11900/7223  & $0 -1100$ &  8871/5474  & \cite{ar87} \cr
\noalign{\vskip 6pt}
SP82 & $0 -1200$ &   9199/5207  & $0 -1100$ &  9103/5283  & \cite{ar83} \cr
\noalign{\vskip 10pt}}
\hrule}}
\vfill
\eject
Table~II. Single-energy (binned) fits of $pp$ data (Pxxx) and combined $pp$
and $np$ data (Cxxx), and $\chi^2$ values.  $\chi^2_E$ is given by the
energy-dependent fit, SM94, and $N_{prm}$ is the number parameters varied in
the fit.
\vskip 10pt
\centerline{
\vbox{\offinterlineskip
\hrule
\hrule
\halign{\hfill#\hfill&\qquad\hfill#\hfill&\qquad\hfill#\hfill
&\qquad\hfill#\hfill&\qquad\hfill#\hfill&\qquad\hfill#\hfill
&\qquad\hfill#\hfill\cr
\noalign{\vskip 6pt} %
Solution&Range~(MeV)&$\chi^2$/$pp$~data&$\chi^2_E$&$\chi^2$/$np$~data&
$\chi^2_E$&$N_{prm}$\cr
\noalign{\vskip 6pt}
\noalign{\hrule}
\noalign{\vskip 10pt}
C  5 & $   4 -   6 $ &   22/28   &   39 &   50/53  &   65 &  6 \cr
\noalign{\vskip 6pt}
C 10 & $   7 -  12 $ &   79/88   &  126 &  134/72  &  189 &  6 \cr
\noalign{\vskip 6pt}
C 15 & $  11 -  19 $ &   17/27   &   45 &  176/213 &  344 &  8 \cr
\noalign{\vskip 6pt}
C 25 & $  19 -  31 $ &  121/114  &  213 &  257/264 &  334 &  8 \cr
\noalign{\vskip 6pt}
C 50 & $  32 -  68 $ &  300/224  &  392 &  616/465 &  684 & 10 \cr
\noalign{\vskip 6pt}
C 75 & $  60 -  90 $ &   46/72   &   53 &  396/311 &  500 & 10 \cr
\noalign{\vskip 6pt}
C100 & $  80 - 120 $ &  136/154  &  177 &  428/344 &  472 & 11 \cr
\noalign{\vskip 6pt}
C150 & $ 125 - 175 $ &  293/287  &  415 &  317/262 &  519 & 13 \cr
\noalign{\vskip 6pt}
C200 & $ 177 - 225 $ &  165/146  &  220 &  605/396 &  697 & 13 \cr
\noalign{\vskip 6pt}
C250 & $ 225 - 275 $ &   66/64   &  146 &  236/220 &  278 & 13 \cr
\noalign{\vskip 6pt}
C300 & $ 276 - 325 $ &  284/256  &  352 &  631/528 &  893 & 17 \cr
\noalign{\vskip 6pt}
C350 & $ 325 - 375 $ &  296/246  &  341 &  496/354 &  664 & 17 \cr
\noalign{\vskip 6pt}
C400 & $ 375 - 425 $ &  556/436  &  648 &  766/552 &  837 & 17 \cr
\noalign{\vskip 6pt}
C450 & $ 425 - 475 $ &  861/647  &  999 &  796/622 &  852 & 18 \cr
\noalign{\vskip 6pt}
C500 & $ 475 - 525 $ & 1378/1067 & 1509 & 1337/851 & 1349 & 18 \cr
\noalign{\vskip 6pt}
C550 & $ 525 - 575 $ &  822/702  &  984 &  620/493 &  695 & 26 \cr
\noalign{\vskip 6pt}
C600 & $ 575 - 625 $ & 1067/703  & 1198 &  425/364 &  575 & 29 \cr
\noalign{\vskip 6pt}
C650 & $ 625 - 675 $ &  859/643  &  860 & 1432/978 & 1727 & 33 \cr
\noalign{\vskip 6pt}
C700 & $ 675 - 725 $ &  809/723  &  851 &  419/407 &  493 & 34 \cr
\noalign{\vskip 6pt}
C750 & $ 725 - 775 $ &  930/768  & 1204 &  508/372 &  621 & 41 \cr
\noalign{\vskip 6pt}
C800 & $ 775 - 825 $ & 1549/1116 & 2096 & 1536/999 & 1633 & 41 \cr
\noalign{\vskip 6pt}
C850 & $ 827 - 875 $ & 1187/882  & 1347 &  380/366 &  421 & 41 \cr
\noalign{\vskip 6pt}
C900 & $ 876 - 925 $ &  310/333  &  434 &  751/628 &  808 & 41 \cr
\noalign{\vskip 6pt}
C950 & $ 926 - 975 $ &  795/623  &  975 &  347/352 &  449 & 41 \cr
\noalign{\vskip 6pt}
C999 & $ 976 -1025 $ &  893/652  & 1064 &  294/331 &  382 & 43 \cr
\noalign{\vskip 6pt}
C110 & $1078 -1125 $ &  705/360  &  835 &  467/326 &  625 & 46 \cr
\noalign{\vskip 6pt}
C125 & $1200 -1296 $ &  890/540  & 1297 &  290/154 &  482 & 48 \cr
\noalign{\vskip 6pt}
P130 & $1261 -1346 $ &  908/583  & 1390 &    0/0   &    0 & 28 \cr
\noalign{\vskip 6pt}
P160 & $1554 -1639 $ &  438/344  &  768 &    0/0   &    0 & 29 \cr
\noalign{\vskip 10pt}}
\hrule}}

\begin{references}
\bibitem{tel}
Those with access to TELNET can run the SAID program with a link to
\hfil\break
VTINTE.PHYS.VT.EDU (128.173.176.61).
The login (password) is: PHYSICS (QUANTUM).  The user may
view the current database and compare our solutions to those of other groups.
\end{references}
\end{document}